\documentclass{article}

\usepackage[preprint]{neurips_2025}

\usepackage[utf8]{inputenc} 
\usepackage[T1]{fontenc}    
\usepackage{hyperref}       
\usepackage{url}            
\usepackage{booktabs}       
\usepackage{amsfonts}       
\usepackage{nicefrac}       
\usepackage{microtype}      
\usepackage{xcolor}         

\usepackage{orcidlink}
\usepackage{longtable}
\usepackage{multirow}

\title{Recognition of Physiological Patterns during
Activities of Daily Living Using Wearable
Biosignal Sensors}

\author{%
  Nicholas Cartocci \orcidlink{0000-0002-4908-9769} \\
  XoLab, Advanced Robotics (ADVR)\\
  Istituto Italiano di Tecnologia (IIT)\\
  Genoa, Italy \\
  \\
  DIBRIS \\
  University of Genoa\\
  Genoa, Italy \\
  \texttt{nicholas.cartocci@iit.it} \\
  \And
  Antonios E. Gkikakis \orcidlink{0002-5339-0431} \\
  XoLab, Advanced Robotics (ADVR)\\
  Istituto Italiano di Tecnologia (IIT)\\
  Genoa, Italy \\
  \texttt{antonios.gkikakis@iit.it} \\
  \And
  Natalia Kurvina \orcidlink{} \\
  DIBRIS \\
  University of Genoa\\
  Genoa, Italy \\
  \texttt{natalia.kurvina@iit.it} \\
  \And
  Natnael Takele \orcidlink{0009-0008-3300-0046} \\
  DIBRIS \\
  University of Genoa\\
  Genoa, Italy \\
  \texttt{natnael.takele@iit.it} \\
  \And
  Fabio Pera \orcidlink{} \\
  Technological Innovation and Safety\\
  INAIL\\
  Rome, Italy \\
  \texttt{f.pera@inail.it} \\
  \And
  Maria Teresa Settino \orcidlink{} \\
  Technological Innovation and Safety\\
  INAIL\\
  Rome, Italy \\
  \texttt{m.settino@inail.it} \\
  \And
  Darwin G. Caldwell \orcidlink{0000-0002-6233-9961} \\
  Advanced Robotics (ADVR)\\
  Istituto Italiano di Tecnologia (IIT)\\
  Genoa, Italy \\
  \texttt{darwin.caldwell@iit.it} \\
  \And
  Jesús Ortiz \orcidlink{0000-0001-9475-1945} \\
  XoLab, Advanced Robotics (ADVR)\\
  Istituto Italiano di Tecnologia (IIT)\\
  Genoa, Italy \\
  \texttt{jesus.ortiz@iit.it} \\
}

\begin{document}

\maketitle

\begin{abstract}
A key aspect of developing fall prevention systems is the early prediction of a fall before it occurs. This paper presents a statistical overview of results obtained by analyzing 22 activities of daily living to recognize physiological patterns and estimate the risk of an imminent fall. The results demonstrate distinctive patterns between high-intensity and low-intensity activity using EMG, ECG, and respiration sensors, also indicating the presence of a proportional trend between movement velocity and muscle activity. These outcomes highlight the potential benefits of using these sensors in the future to direct the development of an activity recognition and risk prediction framework for physiological phenomena that can cause fall injuries.
\end{abstract}

\section{Introduction}

Falling causes severe injuries with significant consequences for the entire healthcare and economic system. According to the World Health Organization (WHO), 684,000 people die, and 172 million suffer disability from falls each year. In addition, the total medical cost resulting from falls of the elderly is \$50 million annually in the US alone (\citet{WHO_2021,Cartocci_2022}). A key aspect of developing fall prevention systems is the early prediction and/or detection of fall events. While a fall detection and ground impact estimation system can rely solely on the use of inertial sensors (\citet{Cartocci_2022,Cartocci_2023,Cartocci_2023_2}), prediction of a loss of balance, heart failure, or fainting can be made solely based on the person's biomedical signals and patterns. To predict the subsequent fall, monitoring the causes (physiological changes) and not the consequences (fall) is necessary. However, this is challenging due to the quantity, quality, and heterogeneity of data and processing time requirements (\citet{Cartocci_2023,Cartocci_2023_2}). The literature lacks data sets based on biomedical and IMU signals that can be useful in falling analysis and used to train and validate algorithms for falling detection and prediction. This paper evaluates these considerations and presents preliminary results obtained by analyzing 22 Activities of Daily Living (ADLs) performed by ten subjects (five females and five males) to recognize physiological patterns and use them in the future to direct the development of a framework to predict the risk of physiological phenomena that may cause fall injures. Once the patterns of ADLs are identified, it is possible to recognize the activity performed by the subject and identify abnormal changes that may be early symptoms of unconsciousness.

\section{Methodology}

A series of experiments with subjects equipped with wearable biosignal sensors was organized to evaluate physiological patterns during Activities of Daily Living. The study was carried out according to the Declaration of Helsinki and approved by the Ethics Committee of Liguria (protocol reference number: CER Liguria 001/2019), 28 October 2019.

\subsection{Activities}

Subjects were required to perform ADLs and dynamic activities during the experiments, but no falls. The primary criterion for selecting dynamic activities was to test the fall prediction algorithms’ robustness, a future work, in tasks involving motions where controlled falls are performed, such as jumping, which do not result in injuries. Table~\ref{tab:list_Activities} describes the 22 activities the subjects participated in, sorted by chronological execution order. During the activity sequence, there are four calibration phases to measure the Maximal Voluntary Contraction (MVC) of the femoral muscles: the subject is asked to push 3 times with the heel up toward a hard surface, such as the base of the treadmill, to measure the MVC of the posterior femoral muscle and to sit on a chair and push 3 times with the leg toward each other with a resistance imposed by an outside person to measure the MVC of the anterior femoral muscle. This procedure is repeated for each foot and is used to evaluate the variation in MVC as the experiments continue and to normalize the muscle data (i.e., EMGs) properly. All experiments were preceded by a 5-minute warm-up session on the treadmill at a constant speed of 3.5 km/h to safeguard the subjects’ health and limit the risk of possible contractures and injuries. The time required to record all trials was approximately 1 - 1.5 hours for each subject.

\subsection{Participants}

For data collection, 10 subjects (five females and five males) were invited to participate. The participants’ ages ranged from 25 to 42 years, their height from 1.55 to 1.91 meters, and their weight from 44 to 83 kilograms, with generally higher male values than female values. The group of participants consists of healthy affiliates of the Istituto Italiano di Tecnologia (IIT).

\subsection{Sensors}

Data were collected with the wearable multi-sensor platform biosignalsplux, PLUX wireless biosignals S.A. (Lisbon, Portugal) using four electromyography (EMG) sensors, an electrocardiography (ECG) sensor, and a piezoelectric respiration (PZT) sensor. The four EMG sensors were placed on the anterior and posterior femoral muscles; the ECG sensor was fixed near the heart, and the PZT sensor was attached frontally under the nipple line. All tests were performed with a sampling frequency of 300 Hz.

\begin{longtable}{llc}
\caption{\small Types of activities carried out during the experiments.\label{tab:list_Activities}}\\ 
\hline
                          & \multicolumn{1}{c}{Activity}                                                                                                   & N. of Repetitions  \endfirsthead 
\hline
Calibrate                 & MAX EMG 80\%, Standing $10s$                                                                                                   & 1                      \\* 
\hline
\multirow{3}{*}{Walking}  & $4km/h$ for $30s$ + $10s$ rest (treadmill)                                                                                               & 1                      \\*
                          & $5km/h$ for $30s$ + $10s$ rest (treadmill)                                                                                                 & 1                      \\*
                          & Walking at own pace                                                                                           & 1                      \\* 
\hline
\multirow{3}{*}{Jogging}  & $6km/h$ for $30s$ + $10s$ rest (treadmill)                                                                                             & 1                      \\*
                          & $7km/h$ for $30s$ + $10s$ rest (treadmill)                                                                                              & 1                      \\*
                          & Jogging at own pace                                                                                           & 1                      \\* 
\hline
\multirow{3}{*}{Running}  & $10km/h$ for $30s$ (treadmill)                                                                                                            & 1                      \\*
                          & $11km/h$ for $30s$ (treadmill)                                                                                                           & 1                      \\*
                          & Running at own pace                                                                                           & 1                      \\ 
\hline
Re-calibrate              & MAX EMG 80\%, Standing $10s$                                                                                                   & 1                      \\* 
\hline
\multirow{3}{*}{Stairs}   & $\sim$2 stairs fast + $10s$ rest                                                                                               & 2                      \\*
                          & $\sim$2 stairs + $10s$ rest                                                                                                     & 2                      \\*
                          & $\sim$2 stairs double steps + $10s$ rest                                                                                       & 2                      \\ 
\hline
Re-calibrate              & MAX EMG 80\%, Standing $10s$                                                                                                   & 1                      \\* 
\hline
\multirow{2}{*}{Ladder}   & \begin{tabular}[c]{@{}l@{}}Ladder climbing (up and down) 5 steps\\ $3s$ stop\end{tabular}                                      & 3                      \\*
                          & \begin{tabular}[c]{@{}l@{}}Ladder climbing (up and down) 5 steps\\ no stop + after the last repetition a jump off\end{tabular} & 3                      \\* 
\hline
\multirow{2}{*}{Jumping}  & Vertical jump                                                                                                                   & 4                      \\*
                          & Leap (2 tiles)                                                                                                                 & 4                      \\* 
\hline
\multirow{4}{*}{Workouts} & Squats                                                                                                                         & 5                      \\*
                          & Jump squats                                                                                                                    & 5                      \\*
                          & Push-ups                                                                                                                       & 5                      \\*
                          & Burpees                                                                                                                        & 5                      \\ 
\hline
Re-calibrate              & MAX EMG 80\%, Standing $10s$                                                                                                   & 1                      \\* 
\hline
\multirow{2}{*}{Sitting/getting up}  &  from a chair                                                                                                & 3                      \\*
                          & from the floor                                                                                              & 3                      \\
\hline
\end{longtable}

\section{Results}

Muscle activation (MA), heart rate (HR), and respiratory rate (RR) were calculated from biomedical signals to assess the variation of these physiological indices according to the activities.

\begin{figure}[htb!]
    \centering
    \includegraphics[width=\textwidth]{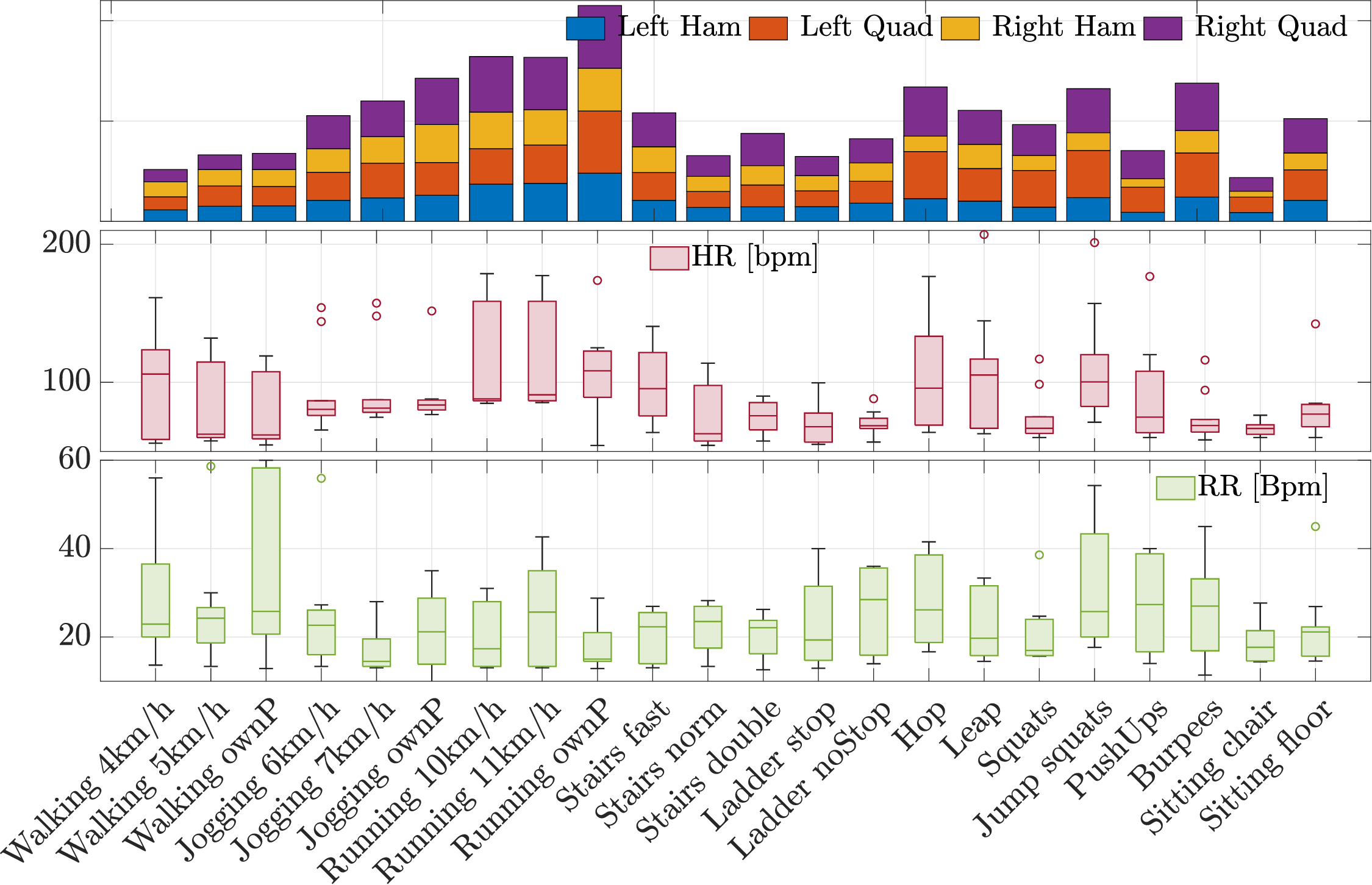}
    \caption{Physiological indices for activities. (a) Above: EMG median value, (b) in the center: Heart Rate (HR), (c) below: Respiration Rate (RR). Ham stands for hamstring, and quad for quadriceps.}
    \label{fig:indices}
\end{figure}

\subsection{Analysis of activity-related EMG signals}

The EMG signals were initially treated with the same process described in (\citet{Altimari_2012}), i.e., rectification and low-pass filtering. Then, the data were normalized with Maximum Voluntary Contraction (MVC) values to make them comparable. In particular, they were divided regarding the 95th percentile of muscle activation during the antecedent calibration phase. After normalizing the data, the median was calculated to evaluate the signal trend. Then, these values were compared for each muscle in which the EMGs were placed for the various activities. The bar graph in Figure~\ref{fig:indices}(a) shows the results obtained considering all subjects. The numerical value on the ordinate has been voluntarily removed to give more meaning to the qualitative differences that can be found.

\subsection{Analysis of activity-related ECG signals}

Heart rate (HR) was estimated using the following frequency method based on ECG signals. Initially, the signals were filtered with a band pass filter with cutoff frequencies of 0.9 Hz (55 bpm) and 3.7 Hz (220 bpm). Subsequently, the filtered signal's Fast Fourier Transform (FFT) was calculated, and the subject's HR during the activity was defined as the sinusoidal frequency with the highest amplitude. Cutoff frequencies were established considering the following considerations presented in the literature. The normal range for resting heart rate is 50–90 beats per minute (bpm), \citet{Spodick1993}, and the maximum heart rate is loosely estimated as 220 minus one's age (\citet{Fox_1972,Riebe_2018}). Figure~\ref{fig:indices}(b) shows the HR thus calculated for the activities studied considering all subjects.

\subsection{Analysis of activity-related respiration band signals}

Similar to the method used to analyze ECG signals and estimate heart rate, respiratory rate (RR) was calculated from the signals of the piezoelectric respiratory band. In this case, the signals were filtered with a band pass filter with cutoff frequencies 0.2 Hz (12 Bpm) and 1 Hz (60 Bpm), given that the regular respiratory rate for an adult at rest is 12 to 18 breaths per minute (Bpm), \citet{Sapra_2023}. The breathing increases to about 40–60 times a minute during exercise (\citet{Troosters_2016}). In Figure~\ref{fig:indices}(c), the RRs calculated for the activities studied considering all the subjects are reported.

\section{Discussion}

By analyzing these indices, it is possible to recognize certain physiological patterns with respect to each activity and to discriminate those heavier and more strenuous, involving, for example, greater muscle activation, increased heart rate, or respiratory rate. Activities that are less strenuous and at lower risk of falling are identified. As the required walking, jogging, and running speed increases, the median for the four muscles seems to increase with a linear profile. Walking at 5 km/h causes a higher average activation than walking at 4 km/h; jogging at 7 km/h causes a higher average activation than jogging at 8 km/h and is similar to running. In addition, the muscle activation of running is greater than that of walking and jogging. In vertical jumping and leaping, it was found that the quadriceps muscles have a higher average activation than the front leg muscles. In contrast, a lower median of muscle activation is observed between subjects in ladder activities than in more explosive activities such as jumping or jumping squats. \\
As previously noted with EMGs, HR increases with the speed imposed on subjects. Furthermore, the highest HR was found for the first activity performed (walking at 4 km/h), running, and jumping. These are also the activities with the most variable HR. Walking and going up and down stairs, as well as stairs, are the activities with the lowest HR, probably also because they are the most carried out by the subjects taken as a sample. \\
The highest RR is found for jumps, squats, and push-ups, while the activities with the lowest are jumping, squatting, and sitting. The results obtained while walking at one’s own pace seem anomalous and present a high variability between subjects. A cause could be found in the frequency with which subjects walk calmly (walking cadence), which could fall within the bandpass filter range (0.2 - 1Hz) and influence the results.

\section{Conclusions and Future Works}

This work introduces a new dataset where 10 subjects of different characteristics are equipped with various biosignal sensors while performing a series of ADL and dynamic activities. The aim is to investigate how biological signals can be used as predictors of falling and fill the gap in the literature that lacks datasets with such signals. The results demonstrate distinctive patterns between high-intensity and low-intensity activities in the heart rate and respiration rate, indicating the presence of distinctive and characteristic patterns and the potential benefits of using such sensors to study a person’s physiological patterns. In addition, the more the speed of movement increases, the more muscle activation increases, following a profile similar to the linear one. These results suggest that these data can also be used in the development of biomedical-based activity recognition. In future work, this dataset will serve as a basis for developing reactive and proactive algorithms for defining and estimating muscular fatigue, cognitive fatigue, and stress levels that could induce a fall, which will result in assessing the risk of an imminent fall before it happens.

\begin{ack}
This research is promoted and conducted in collaboration with the Italian National Institute for Insurance against Accidents at Work (INAIL) under the project ‘Sistemi Cibernetici Collaborativi - Cadute dall’Alto 2 - Tecnologie indossabili per ridurre gli effetti dell’impatto nelle cadute dall’alto’.
\end{ack}


\bibliographystyle{plainnat}
\bibliography{bibl}





\end{document}